\documentclass{aa}

\usepackage{graphics}
\usepackage{times}

\newcommand{\mdf}{\bf}
\newcommand{\diff}{\mathrm d}
\newcommand{\imag}{\mathrm i}

\newcommand{\mincir}{\raise
  -2.truept\hbox{\rlap{\hbox{$\sim$}}\raise5.truept \hbox{$<$}\ }}
\newcommand{\magcir}{\raise
  -2.truept\hbox{\rlap{\hbox{$\sim$}}\raise5.truept \hbox{$>$}\ }}

\newcommand{\eqref}[1]{(\ref{#1})}
\makeatletter
\newcommand{\func}[1]{\mathop {\operator@font #1}}
\makeatother

\def \farcs{\hbox{$.\!\!^{\prime\prime}$}}
\def \farcm{\hbox{$.\!\!^{\prime}$}}

\begin{document}

\thesaurus{02(%
  11.03.4: MS1008.1-1224;  
  12.07.1;  
  12.03.3;  
  12.04.1;} 

   \title{Weak lensing mass reconstruction of MS1008${}.{}$1$-$1224}
   \author{Lombardi M.\inst{1,2}, Rosati P.\inst{1}, Nonino
     M.\inst{3}, Girardi M.\inst{4}, Borgani S.\inst{4,5}, Squires G.\inst{6}}
   \authorrunning{Lombardi M. et al.}
   \offprints{M. Lombardi}
   \mail{lombardi@sns.it}
   \institute{European Southern Observatory,
     Karl-Schwarzschild Stra\ss e 2, 
     D-85748 Garching bei M\"unchen, Germany
     \and
     Scuola Normale Superiore,
     Piazza dei Cavalieri 7, I-56126 Pisa, Italy
     \and
     Osservatorio Astronomico di Trieste, Via Tiepolo 11,
     I-34131 Trieste, Italy
     \and
     Dipartimento di Astronomia, Universit\`a di Trieste,
     via Tiepolo 11, I-34131 Trieste, Italy
     \and
     INFN, Sezione di Trieste, c/o Dipartimento di
     Astronomia, via Tiepolo 11, I-34131 Trieste, Italy
     \and
     Caltech Astronomy M/S 105-24, 1200 E. California Blvd., Pasadena,
     CA, USA}
   \date{Received 22 May 2000; accepted ***date***}

   \maketitle

   \begin{abstract}
     We present an in-depth weak lensing analysis of the cluster
     MS1008${}.{}$1$-$1224 based on deep multicolor imaging obtained
     during the Science Verification of FORS1 at the VLT. The image
     quality (half arcsec seeing) and depth of the VLT images allow
     the shear signal to be mapped with high signal-to-noise and to be
     traced out to $1.2 \, h_{50}^{-1} \mbox{ Mpc}$, near the edge of
     the $6\farcm 8 \times 6\farcm 8$ field of view.  Using BVRI color
     information, as well as 81 redshifts in the field from the CNOC
     survey, background galaxies can be effectively separated from
     cluster and foreground objects. PSF distorsions are found to be
     moderate across the FORS images and thus easily removed. Due to
     the small statistical errors in the mass reconstruction, this
     dataset provides a testing ground where several systematic
     effects (e.g.\ mass-sheet degeneracy, redshift distribution of
     the background sources, cluster galaxy contamination), which are
     involved in the weak lensing analysis, can be quantified.
     Several methods are used to remove the mass-sheet degeneracy
     which is found to dominate the systematic error budget. We
     measure a lower limit to the mass of $2.6 \times 10^{14} \,
     h_{50}^{-1} \mbox{ M}_\odot$ within $1 \, h_{50}^{-1} \mbox{ Mpc}$
     and a ``total'' mass of $5.3\times 10^{14} \, h_{50}^{-1} \mbox{
       M}_\odot$ by fitting a softened isothermal sphere.  We find the
     mass distribution fairly uniform, with no significant
     substructures, in agreement with the virial analysis.  The
     availability of the CNOC redshift data and X-ray observations on
     this cluster allow a comparison of different determinations of
     the mass radial profile. We find the lensing and X-ray
     measurements in excellent agreement, while the mass derived from
     the virial analysis is marginally ($1$--$2 \sigma$) in agreement
     at radii where both methods are reliable.  This analysis
     underscores the importance of systematics in the mass
     determination of clusters, particularly when such a high quality
     dataset is not available or in similar studies at higher
     redshifts.

     \keywords{galaxies:clusters: individual: MS1008.1-1224 -- cosmology:
       gravitational lensing -- cosmology: observations -- cosmology:
       dark matter}     
   \end{abstract}

%

\section{Introduction}

The study of the weak lensing distortion of background galaxies is a
powerful tool for measuring the mass distribution of galaxy clusters.
It has long been recognized (Tyson et al.\ 1984) that the tidal field
of a cluster acts by slightly modifying the images of distant,
background galaxies.  By measuring these small distortions the
projected mass distribution of the lensing cluster can be derived in a
model independent fashion (Kaiser \& Squires 1993), in contrast with
other mass estimators (X-ray measurements of intra-cluster gas density
and temperature, galaxy dynamics).

To date, there are approximately 20 clusters which have been well
studied with lensing techniques (see, e.g., Mellier 1999 for a
review). In several cases the resulting masses are claimed to be
larger (by a factor of two) than those obtained using different
methods. At present, it is not known if this discrepancy is due to
biases and inaccuracies in the X-ray and dynamical estimates, or to
the lensing data reduction method. As a matter of fact, it is not even
clear whether there is a real discrepancy (see Wu et al.\ 1998; see
also Allen 1998).  For example, Boehringer et al.\ (1998) have shown
that different mass estimators are in excellent agreement when very
good quality X-ray and optical data are used.

In this paper, we study the mass distribution of the cluster
MS1008${}.{}$1$-$1224 (in the following, for simplicity, MS1008) using
high quality imaging data obtained during the Science Verification of
FORS1 at the VLT. The depth and seeing quality of these observations in
four different bands (BVRI), make this dataset ideal for weak lensing
mass reconstruction.  The large surface density of background galaxies
in these images, as well as the sharpness of the PSF and its small
variations across the field, significantly improve the accuracy in
measuring the distortion field from object ellipticities.  As a
result, high signal-to-noise shear maps can be obtained. This allows
us to better investigate the effect of systematics involved in the
mass inversion procedure, which in this case exceed statistical
errors. Moreover, the availability of radial velocities for $65$
galaxies in MS1008 from the CNOC survey (Carlberg et al. 1996), as
well as X-ray spectroscopic observations, provide independent
estimates of the total mass.

The paper is organized as follows. In Sect.~2 we introduce the basic
lensing relations and the notation used throughout the paper. The
observations and the imaging dataset are briefly described in Sect.~3.
Section~4 is devoted to the weak lensing analysis of the cluster,
where we describe in detail the mass reconstruction method and the
results obtained. A virial analysis based on the CNOC redshift survey
of MS1008 is provided in Sect.~5. A comparison between the light and
the mass distribution of MS1008, and in particular its mass to light
ratio, are discussed in Sect.~6. Finally, we draw the conclusions of
this study in Sect.~7.

An independent lensing study using the same dataset has recently been
presented by Athreya et al.\ (2000); moreover, the depletion effect on
this cluster has been recently investigated by Mayen \& Soucail (2000,
preprint). We briefly compare our results with these studies in the
conclusions.

Throughout the paper we adopt $\Omega = 0.3$, $\Omega_\Lambda = 0.7$, 
$H_0 = 50\; h_{50} \mbox{ km s}^{-1} \mbox{ Mpc}^{-1}$.

\section{Basic relations}

In this section we briefly review the basic lensing relations. We
mostly follow the notation of Seitz \& Schneider (1997) (see also
Schneider et al.\ 1992).

Define $z_\mathrm d$ to be the redshift of the lens (in this case
$z_\mathrm{d} = 0.30$) and $\Sigma(\vec\theta)$ the two-dimensional
projected mass distribution of the lens, where $\vec\theta$ is a
two-dimensional vector representing a direction on the sky. For a source
at redshift $z$, we define the critical density, $\Sigma_\mathrm c(z)$,
which roughly describes the minimum density that a lens must have in
order to produce multiple images from the source at redshift $z$. This
quantity is given by
\begin{equation}
  \Sigma_\mathrm{c} (z) = \cases{ 
  \infty & for $z \le z_\mathrm d \; ,$ \cr
  \displaystyle\frac{c^2 D(z)}{4 \pi G D(z_\mathrm d) D(z_\mathrm d,
  z)} & otherwise$\; .$ \cr}
\end{equation}
Here $D(z)$ and $D(z_\mathrm d, z)$ are the angular diameter-distance
from the observer to an object at redshift $z$, and from the lens to the
same object respectively. 

The ratio between the lens projected mass density, $\Sigma(\vec\theta)$,
and the critical density, $\Sigma_\mathrm{c}(z)$, is defined as the
dimensionless mass density $\kappa(\vec\theta, z) = \Sigma(\vec\theta) /
\Sigma_\mathrm{c}(z)$. This quantity characterizes the strength of the
lens for sources at redshift $z$; in other words, all the lensing
observables for galaxies at redshift $z$ depend only on
$\kappa(\vec\theta, z)$.

In general, the source galaxies do not lie at a single redshift, $z$,
but are distributed with a some distribution $p(z)$ (here we define
$p(z)$ to be normalized so that $\int p(z) \, \diff z = 1$). This
modifies the mean dimensionless mass density, $\bigl\langle
\kappa(\vec\theta) \bigr\rangle_z$, as
\begin{eqnarray}
  \label{eq:<k>}
  \bigl\langle \kappa(\vec\theta) \bigr\rangle_z &=& \int_0^\infty
  \frac{\Sigma(\vec\theta)}{\Sigma_\mathrm{c}(z)} p(z) \, \diff z
  \nonumber\\
  &=& \frac{4\pi G D(z_\mathrm{d})}{c^2} \Sigma(\vec\theta)
  \int_{z_\mathrm{d}}^\infty \frac{D(z_\mathrm{d}, z)}{D(z)} p(z) \,
  \diff z \; .
\end{eqnarray}
Note that the last equality holds because $\Sigma_\mathrm{c}(z)$ is
defined to be infinity for foreground galaxies.  We define similarly a
mean \textit{shear\/} $\bigl\langle \gamma(\vec\theta) \bigr\rangle_z$
(see, e.g., Seitz \& Schneider 1997 for a definition of the shear
$\gamma(\vec\theta, z)$), which also depends on the last integral of
Eq.~\eqref{eq:<k>}.

 In the weak lensing limit, i.e.\ in the limit $\kappa(\vec\theta, z)
\ll 1$ for all $\vec\theta$ and $z$, all of the statistical lensing
observables depend on $\bigl\langle \kappa(\vec\theta) \bigr\rangle_z$
and $\bigl\langle \gamma(\vec\theta) \bigr\rangle_z$.
 
In this analysis, the method employed to analyze galaxy images and infer
the cluster is based on the weak lensing assumption. As shown by Kaiser
(1999), such an assumption is justified for details on the mass
distribution smaller than $1/\gamma \| \vec\theta \|$.  For the cluster
considered in this paper this limit might not be valid in the central
region, where the presence of a small arc suggests that the cluster is
super-critical or at least nearly critical, and our results in this
region should be interpreted with care.

For simplicity, we will refer in the following to $\kappa(\vec\theta)$
and $\gamma(\vec\theta)$ as averaged quantities, $\bigl\langle
\kappa(\vec\theta) \bigr\rangle_z$ and $\bigl\langle \gamma(\vec\theta)
\bigr\rangle_z$.

\section{Observations}

MS1008 is a rich galaxy cluster drawn from the Einstein Medium
Sensitivity Survey sample (EMSS, Gioia \& Luppino, 1994) at $z=0.302$,
with X-ray luminosity $L_X [0.3-3.5\, \rm{keV}] = 4.5 \times 10^{44}\ 
h_{50}^{-2} \mbox{ erg s}^{-1}$.  MS1008 was also part of the CNOC
Survey (Carlberg et al.\ 1996), whose data we will use extensively
below. This cluster was selected for the Science Verification (SV) of
the first instruments on the VLT-Antu (UT1) to provide a deep
multicolor imaging in B, V, R, I, J and K, \textit{the FORS--ISAAC
  Cluster Deep Field} (see Renzini, 1999).  A detailed investigation
of the mass distribution using gravitational lensing techniques was
one of the main scientific goals of this SV Programme.  Observations
were carried out in January 1999.  FORS (FOcal Reducer/low dispersion
Spectrograph, Appenzeller et al.\ 1998) was used in Standard
Resolution (SR) imaging mode, which provides a $6\farcm8 \times
6\farcm8$ field of view with $0\farcs2$ pixels.  In order to avoid
excessive bleeding and scattered light from two 11 magnitude stars in
the northern part of the field, the FORS MOS mechanism was used in
occulting mode, with two occulting bars on each side of the field
designed to mask out the two bright stars at each position of the
dithering pattern ($10\arcsec$ step). This reduced the field of view
by about $12\%$, which, as discussed below, had a moderate impact on
the weak lensing analysis.

We summarize in Table~1 the main parameters of the optical imaging
data.  Near-IR images obtained with ISAAC, which cover a much smaller
field of view, were not used in the present analysis.

The B-band image was obtained combining all the available frames.  The
VRI-band images utilized in our analysis were obtained coadding only the
best seeing images of the entire set released by ESO, resulting in final
images with measured seeing of $0.5$--$0.6\arcsec$, and depth very
similar to that obtained coadding the whole set.  The flat-fielded
images were registered and combined using procedures described in Nonino
et al.\ (1999). For the coaddition, a modified version of the
\textit{drizzling\/} software originally developed for HST imaging
(Fruchter \& Hook 1999) was used.  The photometric zero points are
accurate at $10\%$ level, and the uncertainty on the zero-points do not
have a significant impact on the lensing analysis.

\begin{table}[!b]
  \caption{Properties of optical images. ``SB limit'' is the
    1-$\sigma$ surface brightness limit; ``seeing'' is the mean FWHM
  obtained by fitting a Moffat profile to a set of bright stars.}
  \small
  \begin{tabular}{l c c c}
    \hline
    \hline
    Band & Exposure & SB limit & Seeing \\
         & (sec.)   & ($\mbox{mag arcsec}^{-2}$) & (arcsec) \\
    \hline
    B    &   $4950$ & $28.25$  & $0.72$ \\            
    V    &   $1350$ & $27.76$  & $0.60$ \\    
    R    &   $1350$ & $27.34$  & $0.52$ \\            
    I    &   $1800$ & $26.48$  & $0.50$ \\    
    \hline
    \hline
  \end{tabular}
\end{table}

\section{Weak lensing analysis}

For the analysis of the weak lensing signal we used the \texttt{IMCAT}
package by Kaiser (Kaiser et al.\ 1995, hereafter KSB) with some
refinements, as described below.

A single master catalog was constructed by performing object detection
on the coadded $\mbox{B} + \mbox{V} + \mbox{R} + \mbox{I}$ image. Weak
lensing analysis was then carried out \textit{independently\/} on the
images in the four filters.  In particular, for each passband, we
measured object shapes and sizes, and performed star/galaxy separation,
PSF correction, and mass reconstruction, as described below.

\subsection{Object detection and shape measurements}

Objects were identified using the \texttt{IMCAT} peak finding algorithm
which utilizes a set of images convolved with top-hat filters of
different sizes (we used filters ranging from $0.5$ to $50$ pixels). For
each object the finding procedure gives, among other parameters, the
coordinates of the center, a detection significance parameter, and an
estimate of the object's size in pixels, which we denote as
$r_\mathrm{g}$.

The local sky level and its gradient were measured for each object by
computing the mode of pixels values on a circular annulus with inner
and outer radii of $16$ and $32$ pixels respectively.  Total fluxes
and half-light radii $r_\mathrm{h}$ were measured on sky subtracted
images using an aperture of radius $3 r_\mathrm{g}$, centered around
each object.

Shape parameters were measured using a Gaussian weight function with
scale length proportional to $r_\mathrm{g}$. The use of a weight
function is needed in order to avoid divergences of the measured
quadrupole moments due to the noise contribution from distant pixels.
The complex ellipticity $\epsilon$ was thus calculated from the
quadrupole moments $Q_{ij}$ using the definition (in the following we
will use both the complex notation and vector notation for quantities
like the ellipticity):
\begin{equation}
  \epsilon = \epsilon_1 + \imag \epsilon_2 = \frac{Q_{11} - Q_{22} + 2
  \imag Q_{12}}{Q_{11} + Q_{22}} \: .
\end{equation}
This quantity describes the object shape regardless of its size.
Objects with large ellipticities are highly elongated along the
direction $\arg(\epsilon) / 2$. The quadrupole moments yield also two
other fundamental quantities: \textit{shear\/} and the
\textit{smear\/} polarizabilities (see below).

Finally, object detections were visually inspected and residual
spurious sources, such as noise, cosmic rays, blends, were removed,
and bright stars were masked out. In addition, objects with low $S/N$
often have $\func{tr}Q <0$ and/or $\det Q <0$ and were excluded.  The
resulting number of objects in the four bands are listed in Table~2
(``good'' detections).

\subsection{Stars/galaxy separation}

Star/galaxy classification was performed independently in each of the
four images.

\begin{figure}[!t]
  \caption{The half-light radius vs.\ magnitude of the objects
    identified on the V band. Crosses represent objects classified as
    unsaturated stars, open squares as galaxies, and dots are
    discarded objects. \label{fig:Vrhmag0}}
\end{figure}

Relatively bright stars can easily be identified on the half light
radius $r_\mathrm{h}$ vs.\ magnitude plot (Fig.~\ref{fig:Vrhmag0}) as
a narrow vertical strip, which turns over for saturated stars. The
star locus is clearly visible against the distribution of galaxies in
the $r_\mathrm{h}$ vs.\ magnitude plot down to $V=23$ (and similarly
$I=22$, $R=22.5$, $B=24$). Thus, more than 60 high signal-to-noise
($S/N$) stars across the field of view (see Table~2) were used to
model the PSF pattern and to correct galaxy ellipticities.  Objects
with $r_\mathrm{h} < r_\mathrm{stars}$ are mainly very faint galaxies
which we excluded in our analysis, as the errors in the determination
of their quadrupole moments are very large. This reduces the catalog
from $\approx 6000$ (``good'' detections in Table~2) to $\approx 3000$
from which background galaxies are selected (see below).

\begin{figure}[!t]
  \centerline{\resizebox{\hsize}{!}{\includegraphics{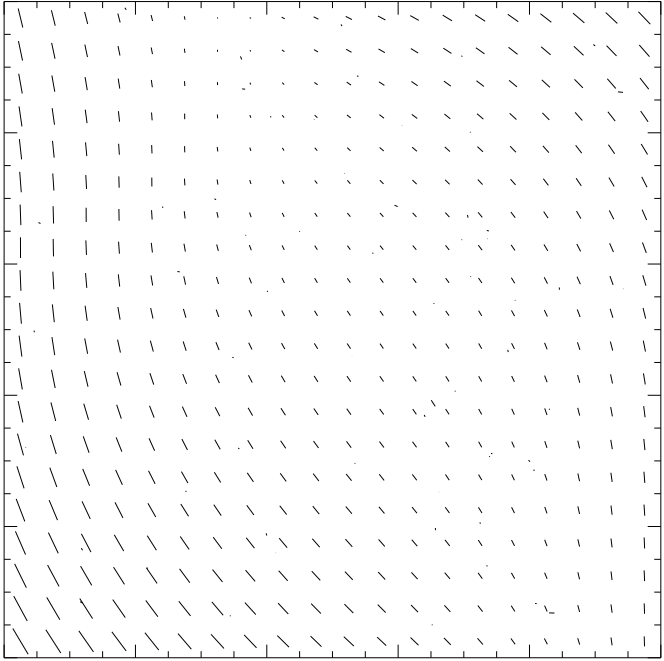}}}
  \caption{The inferred PSF correction for the V image. The segments
    are proportional to the correction and are oriented as the PSF
    (the longest ones corresponding to $4\%$ ellipticities).  Short
    segments outside the grid represent the uncorrected ellipticities
    of stars. \label{fig:Vrhmag2}}
\end{figure}

\begin{figure}[!t]
  \centerline{\resizebox{\hsize}{!}{\includegraphics{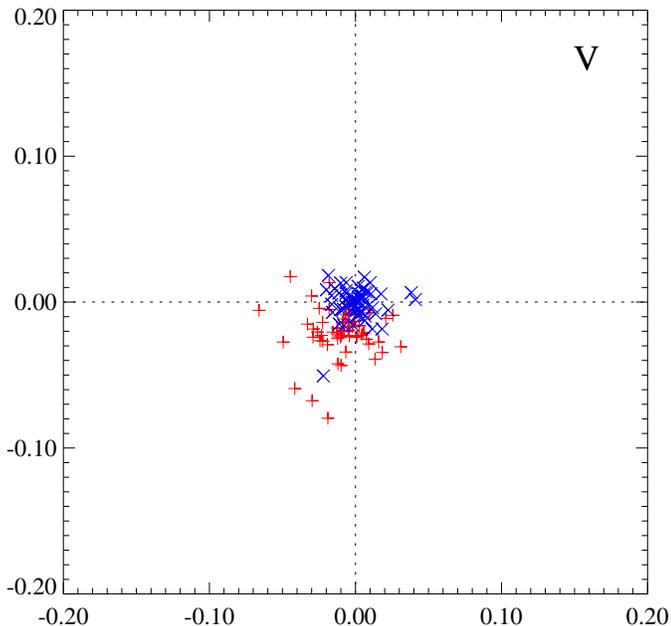}}}
  \caption{The distribution of star ellipticities on the complex plane
    for objects identified on the V image. The ``plus'' signs
    represent the uncorrected ellipticities, while the ``multiply''
    signs represent the ellipticities corrected for the PSF. Similar
    plots are obtained for the other pass-bands. Note that all the
    points are localized near the origin of the plane, i.e. the
    correction is very small for FORS images. \label{fig:Vrhmag1}}
\end{figure}

\begin{table}[!b]
  \caption{The number of detections obtained in the four
    bands. The final number of lensing galaxies used for each band is
    given by the last line (``Relevant galaxies'').}
  \small
  \begin{tabular}{l r r r r}
    \hline
    \hline
                        & \multicolumn{4}{c}{Bands} \\
    Objects             &    B &    V &    R &    I \\
    \hline
    Total detections    & 7969 & 7969 & 7969 & 7969 \\
    ``Good'' detections & 6369 & 6315 & 6213 & 5395 \\
    Stars               &   66 &   60 &   76 &   89 \\
    Galaxies            & 3042 & 3429 & 3584 & 3178 \\
    Cluster members     &  253 &  267 &  269 &  265 \\
    Background galaxies & 2750 & 3125 & 3281 & 2879 \\
    Relevant galaxies   & 1709 & 1844 & 2004 & 1715 \\
    \hline
    CNOC galaxies       & \multicolumn{4}{c}{81 (58 cluster galaxies)} \\
    \hline
    \hline
  \end{tabular}
\end{table}

\subsection{Identification of cluster members}

\begin{figure}[!t]
  \caption{Color-color diagram for the cluster. The open circles
    represent galaxies assumed to belong to the cluster; the filled
    circles represent galaxies with measured redshift in the range
    $0.29$--$0.32$; the dots are the other galaxies; crosses are
    residual galaxies with $R < 21.5$ (see Fig.~\ref{fig:colormag}).
    \label{fig:colorcolor}}
\end{figure}

\begin{figure}[!t]
  \caption{Color-magnitude diagram for the cluster. 
  Symbols are as in Fig.~\ref{fig:colorcolor}. \label{fig:colormag}}
\end{figure}

The deep multicolor photometry provided by the FORS observations, in
combination with a large number of spectroscopic redshifts on MS1008
available from the CNOC survey (Yee et al.\ 1998), allows the cluster
galaxies to be effectively separated from the foreground and
background galaxy population.  Of the 81 galaxies with measured
redshifts in the FORS field, there are 58 cluster galaxies with
$0.29<z<0.32$, which lie along a distinctive ``finger'' of galaxies in
the color-color plot (Fig.~\ref{fig:colorcolor}). This clump in the
$V-I$ vs.\ $B-R$ includes most of the reddest early type galaxies in
the cluster (with $V-I\approx 1.8$), as well as bluer cluster members,
which are not simply distinguishable in a color-magnitude plot. Fewer
late type cluster galaxies extend into a region dominated by
foreground and background galaxies (filled dots with $V-I < 1.4$).
Thus, we are confident that the ``finger tip'' in
Fig.~\ref{fig:colorcolor} encompasses the majority of the cluster
galaxies (254), 23\% of which are spectroscopic members.

The color-mag diagram in Fig.~\ref{fig:colormag} shows the same
selected objects, mostly lying along the cluster red sequence. A
further magnitude cut is shown, flagging all the remaining objects
brighter than $R = 21.5$ as likely cluster or foreground galaxies
(crosses in Fig.~\ref{fig:colorcolor} and \ref{fig:colormag}).
Approximately 3000 objects were thus selected as ``background
galaxies''.

\subsection{Determination of the Shear}

For gravitational lensing analyses, there have been several
investigations to determine the optimal method for measuring galaxy
shapes, correcting for non-gravitational shape distortions, and
calibrating the gravitational shear estimates from the galaxy shape
data (e.g., KSB; Luppino \& Kaiser 1997; Kuijken 1999; Kaiser 2000).
For ground based data, the standard approach adopted in most of the
literature follows the KSB formalism to correct for PSF anisotropies
and the Luppino/Kaiser (LK) algorithm for calibrating losses due to
seeing and pixelization.  We adopted this procedure, which,
qualitatively, determines the shear map correcting for the isotropic
and anisotropic part of the PSF, using a set of high $S/N$ stars in
the field.

Having adopted the method for determining the shear, there are still
several technical issues in the implementation of the algorithm that
can introduce a bias into the results. As an attempt to minimize such
effects (which we describe in more detail below), we performed a
double-blind lensing analysis: two of the authors did independent,
complete shear analyses, using custom designed software
implementations, and somewhat different estimators to calibrate the
effects of seeing and PSF anisotropies. The final results from both
studies were in agreement, we thus present the results from only one
implementation below. The second approach is described in the
Appendix.

Defining $\epsilon^\mathrm{s}$ to be the \textit{unlensed, source\/}
ellipticity, we write the observed ellipticity as
\begin{equation}
  \label{eq:e-e^s}
  \epsilon_i \simeq \epsilon^\mathrm{s}_i + P^\gamma_{ij} \gamma_j +
  P^\mathrm{sm}_{ij} p_j \; ,
\end{equation}
where we have used the Einstein convention on repeated indexes. This
equation simply states that, to the first order in $\gamma$ (weak
lensing limit), the observed ellipticity depends linearly on the shear
$\gamma$. The quantity $P^\gamma$ is called \textit{pre-seeing shear
  polarizability}, and is given by
\begin{equation}
  P^\gamma_{ij} = P^\mathrm{sh}_{ij} - P^\mathrm{sm}_{ij}
  \left\langle \frac{P^\mathrm{sh}_{kk}}{P^\mathrm{sm}_{kk}}
  \right\rangle_* \; .
\end{equation}
Here $P^\mathrm{sh}$ and $P^\mathrm{sm}$ are, respectively, the
\textit{post-seeing shear polarizability\/} and the \textit{smear
  polarizability\/}; both quantity are calculated by \texttt{IMCAT}
using the object quadrupole moments. 
The symbol $\langle \, \cdot \,
\rangle_*$ indicates that the average is taken over the stars.  
We refer to Luppino
\& Kaiser (1997) for further details.

The above relation was applied as follows. We calculated for each
detected star the quantity $P^\mathrm{sh}_{kk} / P^\mathrm{sm}_{kk}$,
and fitted the derived values with a second order polynomial. The
quantity $P^\mathrm{sh}_{kk} / P^\mathrm{sm}_{kk}$ gives information on
the isotropic part of the PSF, and, as expected, we noticed little
variation of this quantity across the field.

The anisotropic part of the PSF was obtained in a simple way by applying
Eq.~\eqref{eq:e-e^s} to our set of stars. For stars, $\gamma = 0$
(because there is no lensing effect), and $\epsilon^\mathrm{s} = 0$
(images of stars would be perfectly round without PSF anisotropies). In
conclusion, we have $p = (P^\mathrm{sm})^{-1} \epsilon$. As before, in
order to obtain a smooth map of the anisotropies, we have fitted the $p$
with a second order polynomial, $p(\vec\theta)$, which is used in
Eq.~\eqref{eq:e-e^s} to perform the correction. Figure~\ref{fig:Vrhmag2}
shows the fitted polynomial $p(\vec\theta)$ in the V band; the plots of
the original and corrected ellipticities of stars are given in
Fig.~\ref{fig:Vrhmag1} (the plots for the other bands are very similar).

The determination of the shear map was handled with special care.
First, galaxies characterized by a small size, a low value of the
detection significance $\nu$, or a small $P^\gamma$ have been excluded
from the catalog. For these galaxies errors on the determination of
the shape parameters are quite large; moreover, they are significantly
affected by the isotropic part of the PSF and thus they do not
contribute to the lensing signal but only introduce noise.  For the
remaining galaxies we have adopted a \textit{weighted median\/}
defined in the following way. For each galaxy, we have evaluated an
``individual shear estimator'' given by
\begin{equation}
  \label{eq:shear}
  \gamma_{ij}^{(n)} = \bigl\langle
  \bigl(P^{\gamma(n)}\bigr)^{-1}_{iji'j'} \bigl(
  \chi_{i'j'}^{\mathrm{obs}(n)} - P^{\mathrm{sm}(n)}_{i'j'i''j''} p_{i''j''}
  \bigr) \bigr\rangle \; ,
\end{equation}
Then we assign, for each point $\vec\theta$ of the field of
observation $\Omega$, a weight $W^{(n)}(\vec \theta) = W(\vec\theta,
\vec\theta^{(n)})$ to each galaxy. Finally, we evaluate for each point
$\vec\theta$ of the field a weighted median of both
$\gamma_1(\vec\theta)$ and $\gamma_2(\vec\theta)$. [The weighted
median $\tilde x$ of a set of $N$ real values $x_n$ with weights $w_n$
is defined as the number such that the sum of weights on $n$ such that
values $x_n < \tilde x$ is equal to the sum of weights on $n$ such
that $x_n > \tilde x$. In other words, we must have
\begin{equation}
  \label{eq:4}
  \sum_n w_n \func{sgn}(x_n - \tilde x) = 0 \; ,
\end{equation}
where $\func{sgn}(x) = x / |x|$ is the function ``sign'' defined to be
zero for $x = 0$. Since in general the condition \eqref{eq:4} cannot
be exactly satisfied, a linear interpolation is used.]\@
Alternatively, one could use a simple weighted average. Simulations,
however, have shown that the median behaves much better. This is due
essentially to a small number of galaxies with very poor shape
determinations.  In such a situation, clearly, the average is not a
\textit{robust\/} shear estimator, while the median is. As an
alternative solution, one could use a suitable supplementary weight to
take into account the (estimated) error on the determination of the
shape parameters of each galaxy.  This solution is \textit{not easy\/}
to implement because it is not always obvious if the shape of a galaxy
has been measured correctly or not. Moreover, the use of supplementary
weights require a fine tuning of the weighting method and this clearly
leads to some arbitrariness on the reconstruction. We note that a
straightforward application of the mean (standard practice in the
literature) instead of the median introduces in the reconstructed
shear map spurious ``substructures'' (i.e.\ secondary maxima). This is
merely due to small, faint sources for which the measurement of the
ellipticity is unreliable.

For the reconstructions shown below the weight $W(\vec\theta,
\vec\theta')$ has been implemented using a Gaussian of argument $\|
\vec\theta - \vec\theta' \|$ and scale length $\sigma_W = 30''$
($\vec\theta'$ is the galaxy position). As explained by Lombardi \&
Bertin (1998a, 1998b), the measured shear map has noise properties
that strictly depend on the weight function used. In particular, we
obviously cannot expect to detect on the shear map details smaller
than the characteristic scale of the spatial weight function. For the
reconstructions shown here we used a Gaussian spatial weight function
with a scale length $\sigma_W$.

\subsection{Cluster mass}

The dimensionless projected mass map, $\kappa(\vec\theta)$, was
obtained using an optimized reconstruction algorithm (Seitz \&
Schneider 1996; Lombardi \& Bertin 1998b; Lombardi 2000).

Along with $\kappa(\vec\theta)$, we also calculated
$\sigma_\kappa(\vec\theta)$, an estimate of the error on the
reconstructed mass distribution. As described in Lombardi \& Bertin
(1998b), this estimate takes into account the intrinsic statistical
error of weak lensing reconstructions, basically proportional to
$\bigl\langle |\epsilon^\mathrm{s}|^2 \bigr\rangle$. We stress that
$\sigma_\kappa(\vec\theta)$ is a \textit{local\/} estimate of the
error; correlation on the scale of the weighting function is expected
on the reconstructed map.

In order to obtain the cluster mass in physical units we need to
estimate the redshift distribution $p(z)$ of the background galaxies
for each passband.

\begin{figure}[!t]
  \centerline{\resizebox{\hsize}{!}{\includegraphics{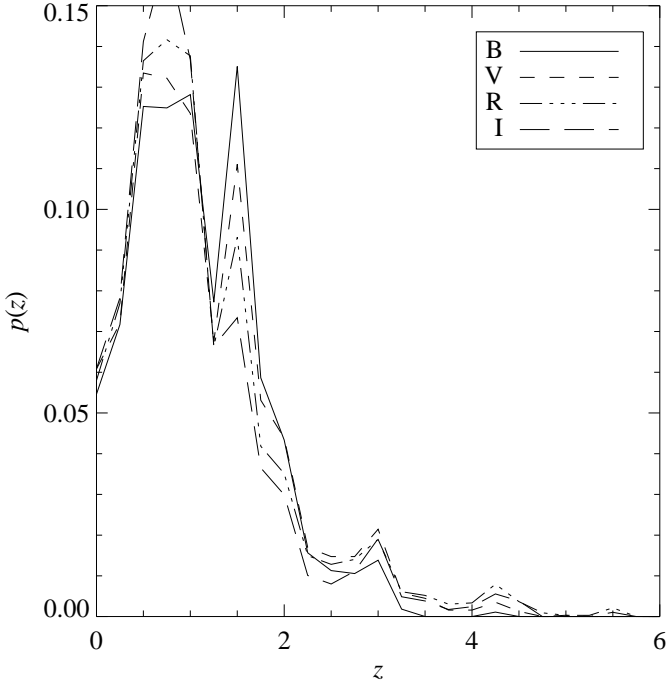}}}
  \caption{The  redshift distribution $p(z)$ of the background
    galaxies in the B, V, R, and I filters derived by resampling the
    catalog of photometric redshift in the HDF-N.
\label{fig:pz}}
\end{figure}

A \textit{statistical\/} estimate of $p(z)$ in each magnitude bin can
be obtained by resampling the catalogue of photometric redshifts of
Fern\'andez-Soto et al.\ (1999) in the Hubble Deep Field North
(HDF-N).  The redshift distribution of the HDF-N derived from
photometric methods is believed to be fairly reliable, given the large
number of spectroscopic redshifts available in the field and the good
photometric accuracy of the HDF-N images.  The drawback in using the
HDF-N as a reference field is of course that, for such a small field,
$p(z)$ might not be representative. However, this is currently one of
the few fields from which $p(z)$ can be estimated down to our
magnitude limits in an empirical fashion, without resorting to galaxy
evolution models for $N(z,m)$.  In our analysis, we assumed that the
redshift distribution of background field galaxies, as a function of
magnitude, matches the one in the HDF-N. We neglected the
magnification bias, i.e.\ the brightening of the background galaxies
due to gravitational lensing, which should not be substantial in
MS1008 given the lack of prominent strong lensing features.

If $N_\mathrm{b}(i)$ is the observed number of background galaxies in
our image, in the $i$-th magnitude bin, and $N$ the total number of
galaxies detected in the image, we can write
\begin{equation}
  \label{eq:p(z)}
  p(z) = \sum_i \frac{N_\mathrm{b}(i)}{N} p_\mathrm{ref}(z; i) \; .
\end{equation}
Here $p_\mathrm{ref}(z; i)$ is the redshift distribution in the
reference field, in the $i$-th magnitude bin, which
can be written as
\begin{equation}
  p_\mathrm{ref}(z; i) = \frac{1}{N_\mathrm{ref}(i)} \sum_n
  \delta\bigl(z - z^{(n)} \bigr) \; .
\end{equation}
$N_\mathrm{ref}(i)$ is the total number of galaxies in the reference
field in the $i$-th magnitude bin, and $z^{(n)}$ is the (photometric)
redshift of the $n$-th galaxy in the reference field (the sum running
on all galaxies $n$ found in a given magnitude bin).

\begin{figure}[!t]
  \centerline{\resizebox{\hsize}{!}{%
      \includegraphics{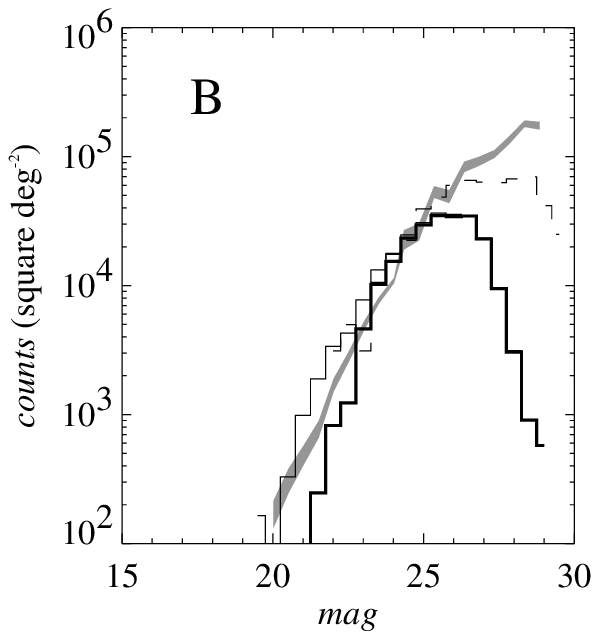}%
      \includegraphics{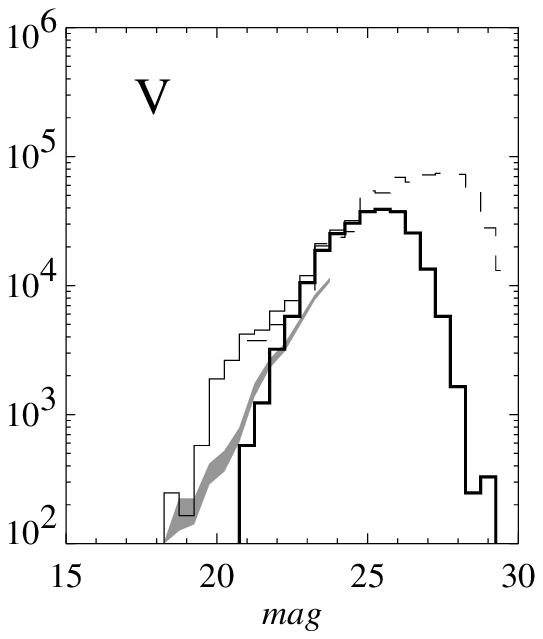}}}
  \centerline{\resizebox{\hsize}{!}{%
      \includegraphics{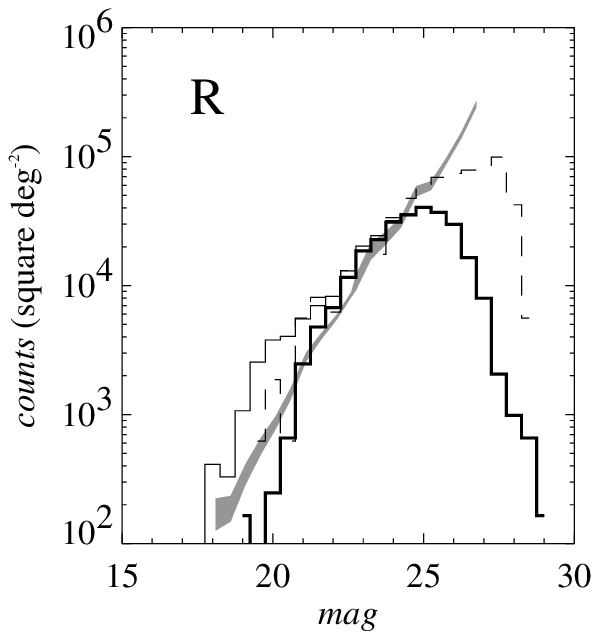}%
      \includegraphics{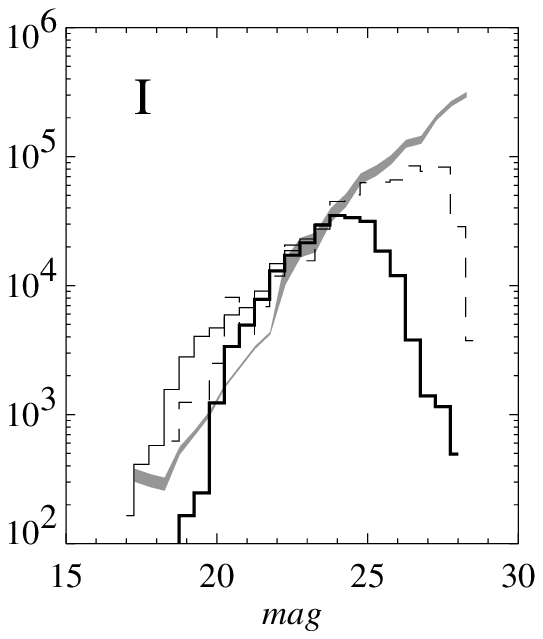}}}
  \caption{Observed number counts (solid line) for the galaxies in the
    four bands. The upper and lower solid lines correspond,
    respectively, to the number counts of all galaxies in the FORS
    field and those identified as background galaxies. We have
    superimposed (dashed line) the number counts of the photometric
    redshift catalog in the HDF-N reference field (Fern\'andez-Soto
    et al.\ 1999).  The shaded areas are number counts (with
    statistical errors) from the literature (Arnouts et al.\ 1997;
    Williams et al.\ 1996).\label{fig:B-Inc}}
\end{figure}

Figure~\ref{fig:pz} shows the redshift distribution for the background
galaxies, $p(z)$, derived with this method.  A comparison of the
observed number counts for our field with those in HDF-N is given in
Fig.~\ref{fig:B-Inc}. The thick solid histogram represents the number
counts of background galaxies, $N_\mathrm{b}(i)$.  The overall number counts
(thin solid histogram) clearly shows the excess at bright magnitudes
due to cluster galaxies. The number counts of HDF-N galaxies with
photometric redshifts (dashed curve) represent $N_\mathrm{ref}(i)$ in
the above notation.

The knowledge of $p(z; i) \equiv p_\mathrm{ref}(z; i)$ for the
background galaxies allows us to evaluate the projected mass
distribution with an \textit{optimized\/} method, which has the
advantage of minimizing the errors on the mass distribution by
properly weighting the contribution of each magnitude bin to the
density $\Sigma(\vec\theta)$.

For each magnitude bin $i$, we can define an \textit{effective\/}
critical density as $\Sigma(z_{\mathrm{eff}, i}) = \bigl\langle
\Sigma^{-1}_\mathrm{c}(z) \bigr\rangle^{-1}$, where the average is
evaluated using the redshift probability distribution $p(z; i)$ (bins
of $1$ magnitude were used).  By using a weight proportional to
$1/\Sigma_\mathrm{c}^2(z_{\mathrm{eff}, i})$ in the average of
Eq.~\eqref{eq:shear}, we take into account the fact that the lensing
signal measured for galaxies in the bin $i$ decreases as the effective
critical density increases.

It is well known that the so-called \textit{mass-sheet invariance\/}
ultimately limits the accuracy in the determination of the
\textit{total\/} mass of the cluster, so that only a lower limit is
strictly provided by the weak lensing analysis.  

A popular method to remove the mass-sheet degeneracy is to assume that
the cluster mass density vanishes at large distance from the cluster
center.  In doing this, we computed the radial profile of the mass map
(see below) and set the \textit{minimum\/} of this profile to zero.
In this way, we almost certainly underestimate the mass density
because the field of view is not large enough to include distant
regions of the cluster.  We also note that taking the minimum of the
two-dimensional mass map is generally more risky.  In fact, because of
the noise, the measured mass map is not bound to be positive
everywhere; on the other hand, the expected noise in the radial
profile is significantly smaller (see shaded area in
Fig.~\ref{fig29.ps}) and can be ignored for the purpose of removing
the sheet degeneracy.

Another method which is often used is to assume some particular model
for the projected cluster mass distribution, and to remove the sheet
degeneracy by requiring that the observed mass matches the model.  In
the case of a singular isothermal sphere the correction to perform for
the measured integral radial profile can be written explicitly.  If
the mass-sheet degeneracy in the observed radial profile
$M_\mathrm{obs}(r)$ has been removed by assuming that the mass
vanishes at a given radius, $r_\mathrm{ap}$, i.e.\ by setting
$\bigl\langle \kappa(r_\mathrm{ap}) \bigr\rangle = 0$, then the
corrected mass profile $M_\mathrm{corr}(r)$ is given by
\begin{equation}
  \label{eq:1}
  M_\mathrm{corr}(r) = \frac{r + r_\mathrm{ap}}{r_\mathrm{ap}}
  M_\mathrm{obs}(r) \; .
\end{equation}
Note, in particular, that when $r = r_\mathrm{ap}$ the mass is
underestimated by a factor $2$.
{\mdf In our mass estimates we have used both techniques described above.}

\subsubsection{Projected mass distribution and radial profiles}

\begin{figure*}[!t]
  \centerline{\resizebox{\hsize}{!}{%
      \includegraphics{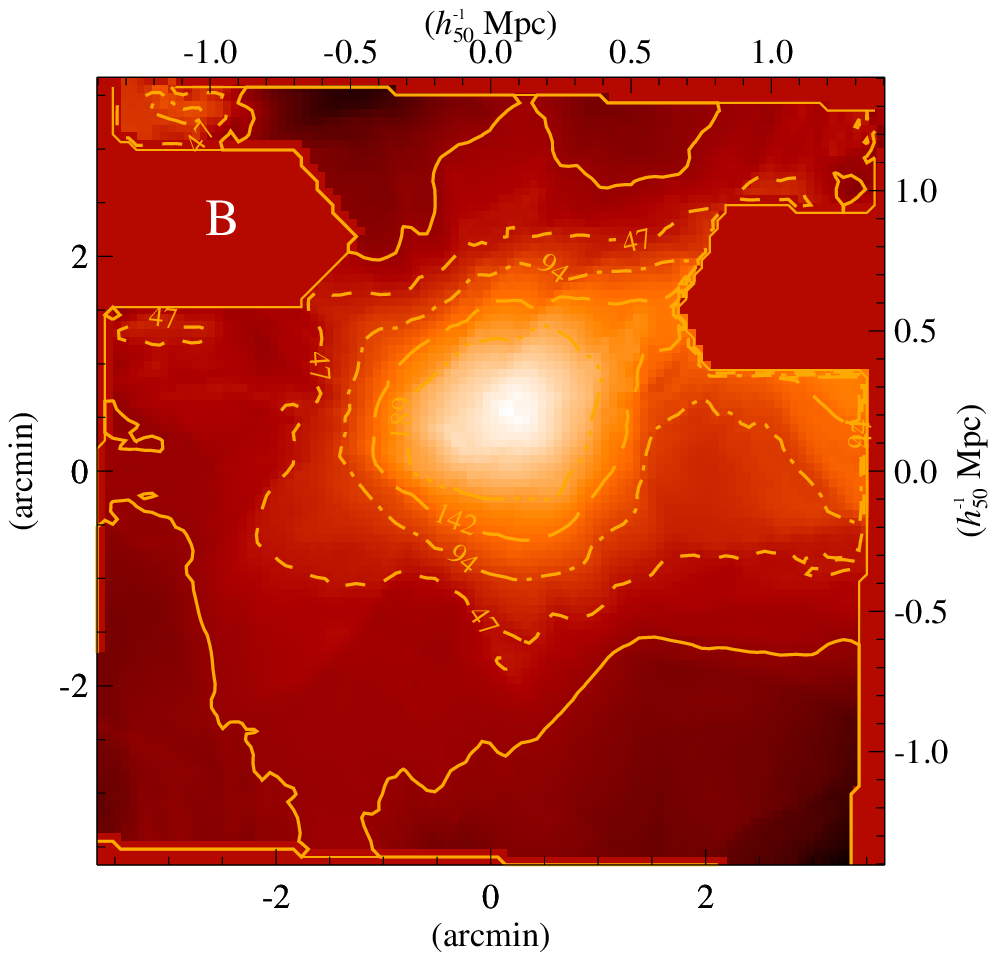}%
      \includegraphics{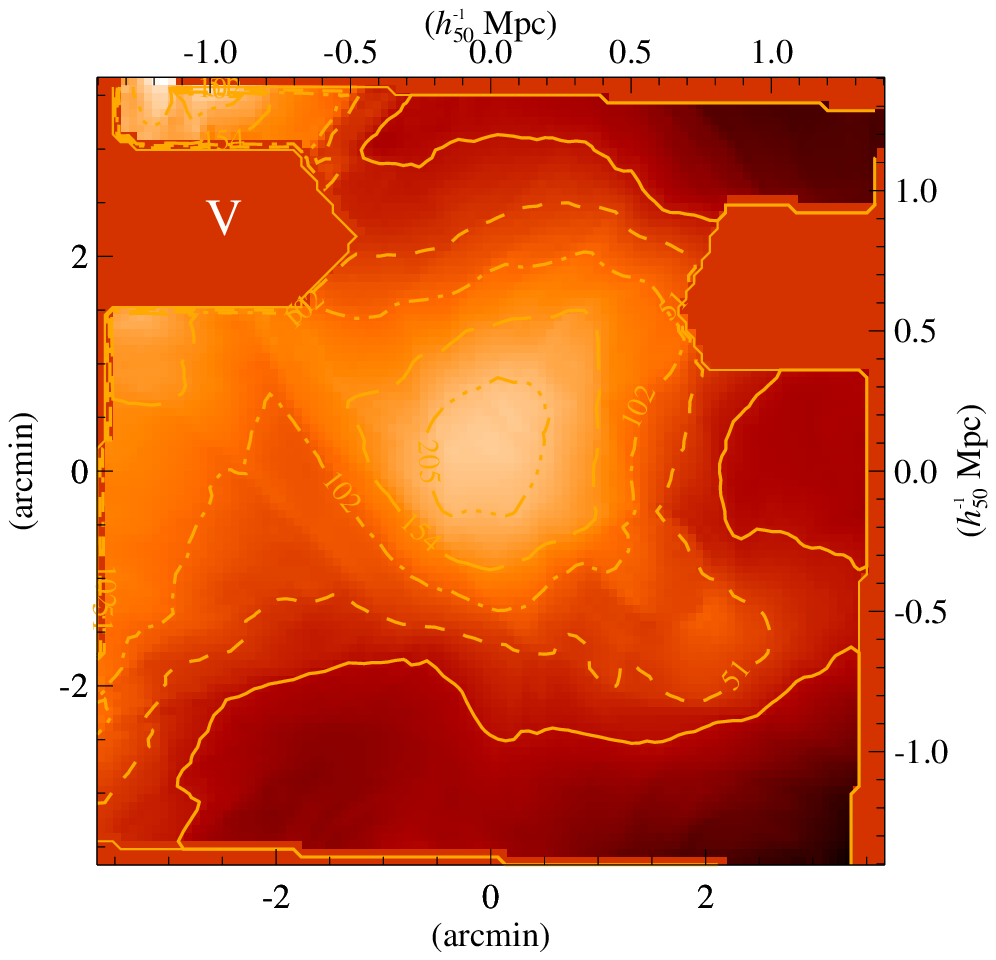}}}
  \centerline{\resizebox{\hsize}{!}{%
      \includegraphics{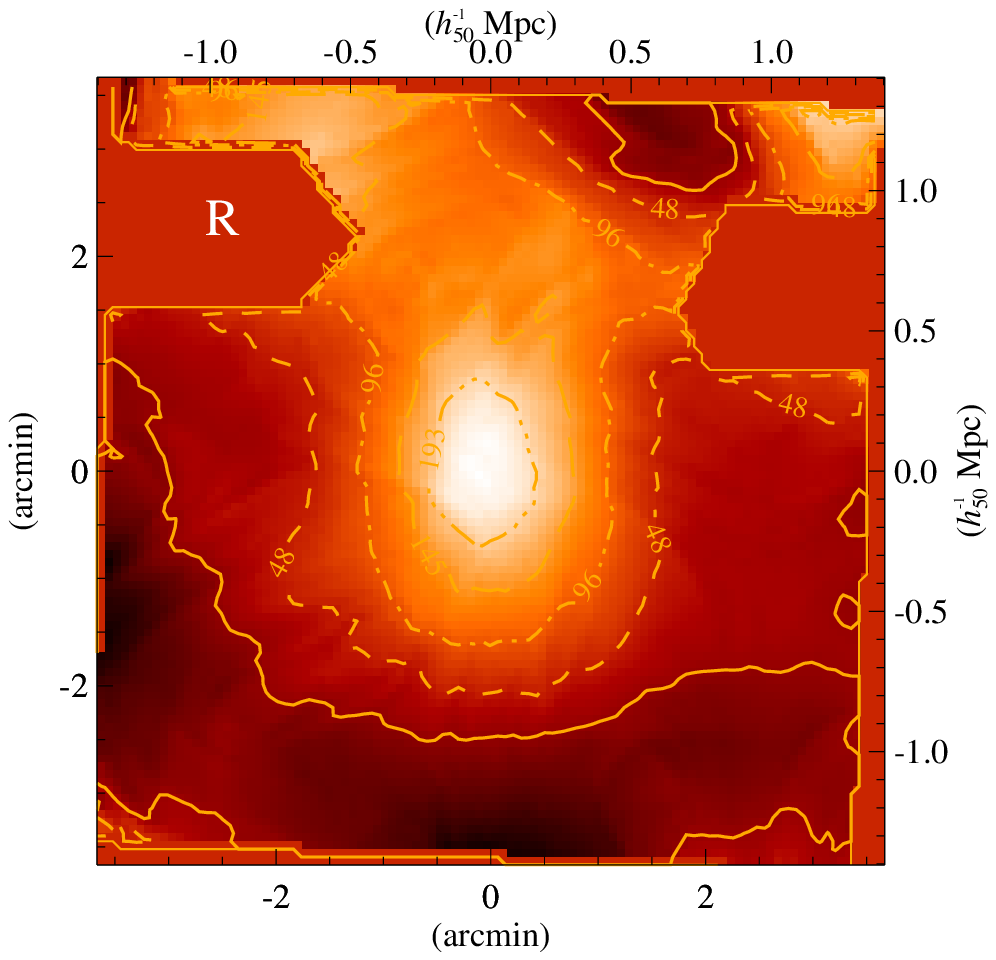}%
      \includegraphics{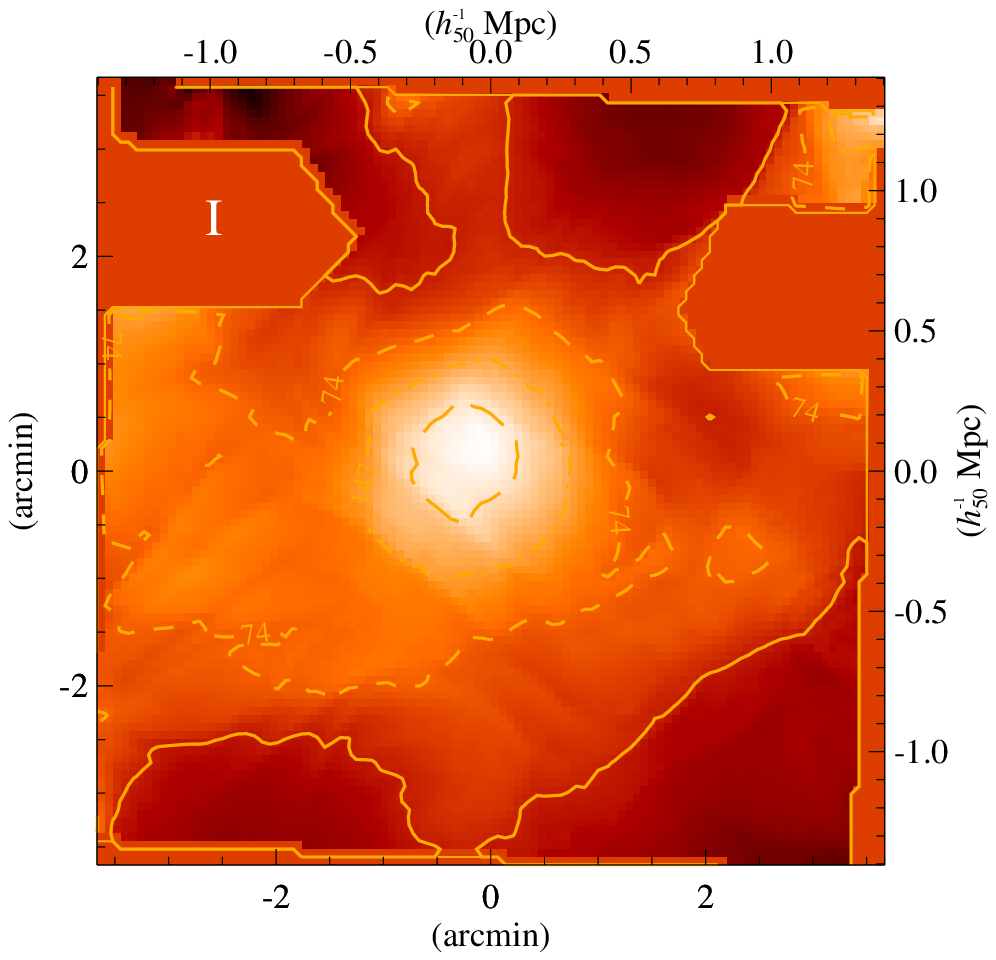}}}
  \caption{The mass distribution obtained from the
    four bands in units of $\mbox{M}_\odot \ h_{50} \mbox{ pc}^{-2}$.
    The solid line represent the contour of zero mass density. Top
    Left: B band; Top Right: V band; Bottom Left: R band; Bottom
    Right: I band.\label{fig:B-I}}
\end{figure*}

The mass maps obtained in the four filters are shown in
Fig.~\ref{fig:B-I} in physical units.  The average mass map 
{\mdf (i.e., the arithmetic mean of the four independent mass maps)}
is also shown in Fig.~\ref{fig:mass}, superimposed on the V band image
of the cluster.  The peak of the projected mass appears to be
displaced $\sim\!  30\arcsec$ north of the cD galaxy. A significant
extension of the core to the north is also apparent. We should
emphasize that, although the mass reconstruction is performed
independently in the four images, the shear maps are not completely
independent because we basically are using the same set of galaxies.
In other words, we expect that the error on the averaged mass map is
smaller than the error in the single bands (because we are reducing
errors on the ellipticities measurements), but by a factor less than
two (because we use the same galaxies).

\begin{figure*}[!t]
  \vskip 0.2cm
  \caption{The mass distribution, obtained by simply averaging the mass maps
    in figure~\protect\ref{fig:B-I}, superimposed on the V image of
    MS1008.  Dashed contours are labeled in units of $\mbox{M}_\odot \ 
    h_{50} \mbox{ pc}^{-2}$. Solid contours correspond to zero mass
    density. The two blank regions on the top correspond to two bright
    stars masked out during the observations.\label{fig:mass}}
\end{figure*}

\begin{figure*}[!t]
  \centerline{\resizebox{\hsize}{!}{%
      \includegraphics{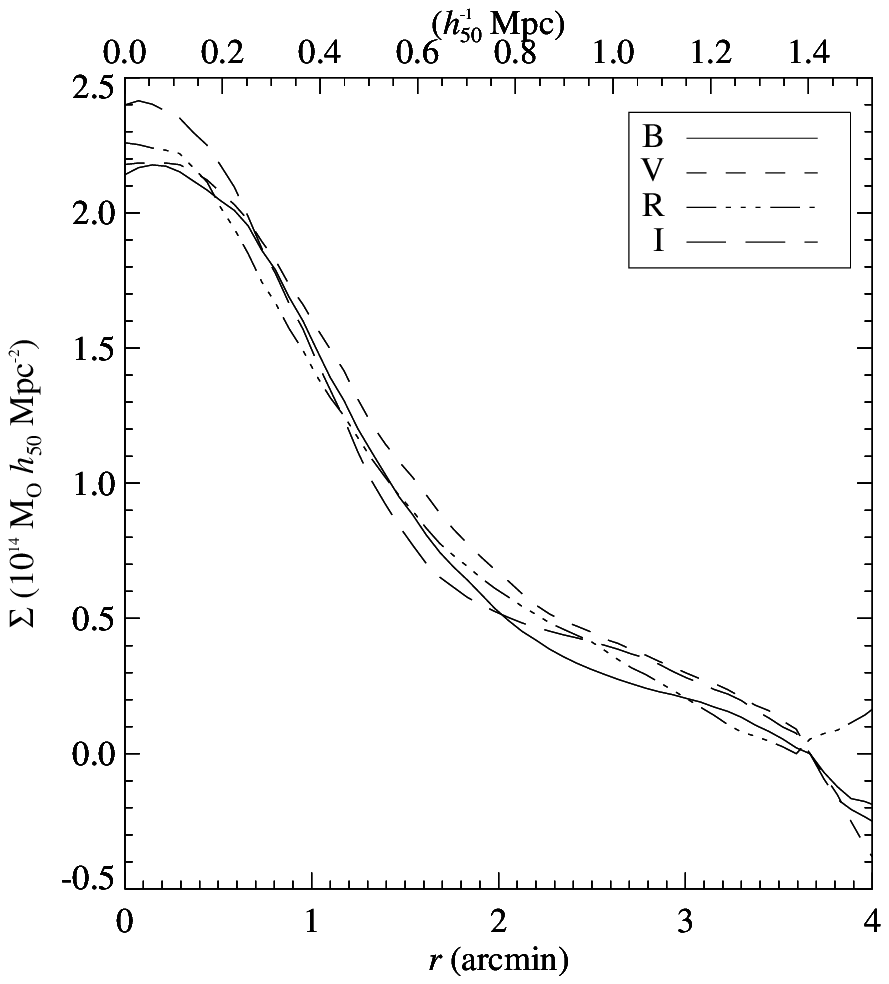}%
      \includegraphics{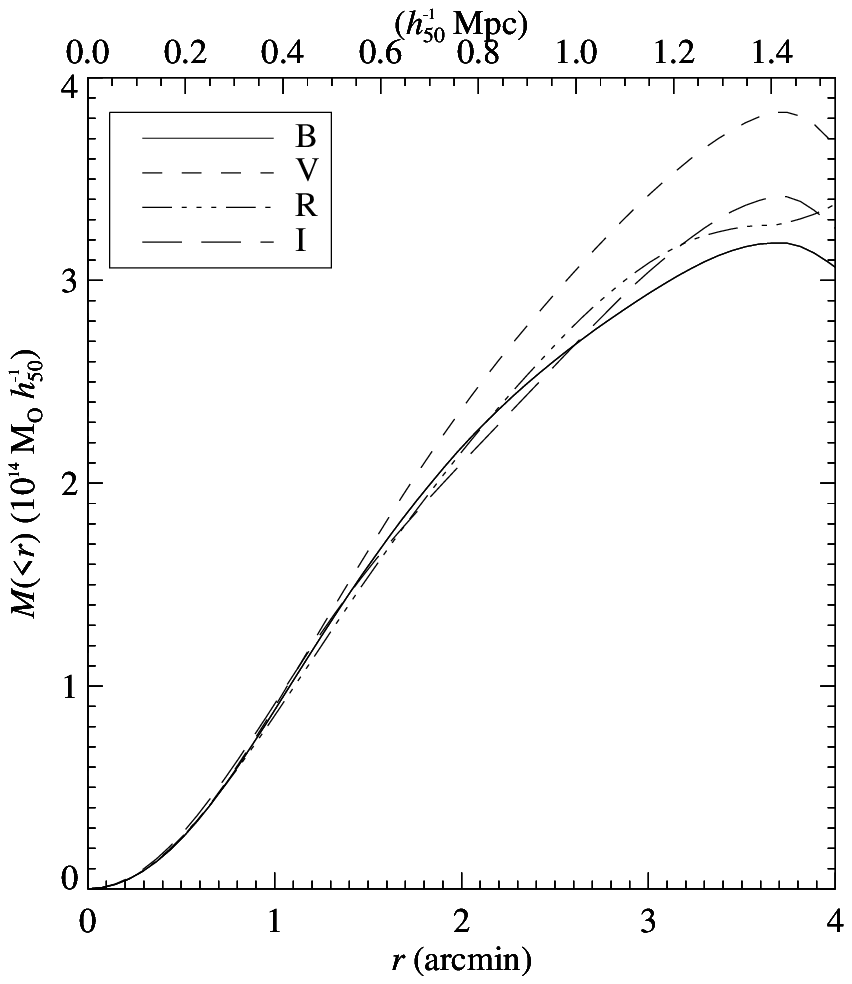}}}
  \caption{The radial profile (left) and the cumulative radial profile
    (right) of the mass distributions obtained in the four
    bands.\label{fig:fig17-18}}
\end{figure*}

The radial profiles shown in Fig.~\ref{fig:fig17-18} were obtained by
averaging the mass distributions on annuli of increasing radii. The
same center was used for all bands, corresponding to the maximum
of the lensing signal.

The azimuthal average was weighted taking into account the local
estimate of the error for the lens mass distribution, which depends on
the local number of galaxies and on the smoothing function used. The
use of a weighted average is crucial at larger radii, where two large
regions are masked out.  The variance $\sigma_\kappa^2(\vec\theta)$
was also used to estimate the errors on the radial profiles.  These
errors should be taken with caution, since different points of the
mass profile are correlated.  The differential radial profiles
(Fig.~\ref{fig:fig17-18}, left) are in excellent agreement with each
other, and their scatter is smaller than $5\%$. This can also be taken
as a rough estimate of the error in the mass density. The cluster mass
as function of radius can be read off the cumulative radial profile in
Fig.~\ref{fig:fig17-18} (right panel).  
{\mdf In order to make these cumulative mass profiles we have broken
  the mass-sheet degeneracy by assuming, for each passband, that the
  (differential) radial profile vanishes at large radii (more
  precisely, at $r = 3' \, 40''$, corresponding to our field of view).}
A comparison of different determinations of $M(<r)$ in the four
passbands also gives a quantitative estimate of the systematics
involved in the measurement of the total mass, in particular those
associated with the way the mass-sheet degeneracy has been removed
(which is responsible for the larger scatter observed in the
cumulative profiles with respect to the differential radial profiles).
For example, the mass out to $1 h_{50}^{-1}$ Mpc is in the range
$(2.7\div 3.0)10^{14} \mbox{ M}_\odot$, i.e. $\sim\!  10\%$.

\begin{figure}[!t]
  \centerline{\resizebox{\hsize}{!}{%
      \includegraphics{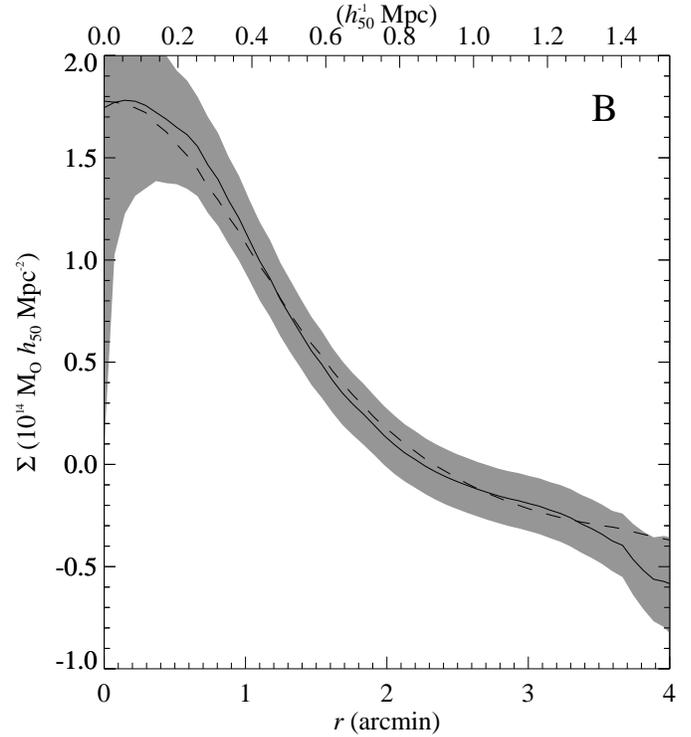}}}
  \caption{The radial mass profile obtained in the B filter
    with the errors band. The dashed curve is the result of a fit with
    a smoothed isothermal sphere. Very similar results are obtained
    for the other filters.\label{fig29.ps}}
\end{figure}

The radial mass profiles can also be fitted with a softened isothermal
model (e.g Schneider et al.\ 1992). This model is
characterized by the core radius $r_\mathrm{c}$ and the central
density $\Sigma_0$; the projected mass distribution is then
\begin{equation}
  \label{eq:iss}
  \Sigma(\vec \theta) = \Sigma_0 \frac{1 + \| \vec\theta/r_\mathrm{c}
  \|^2 / 2}{\bigl( 1 + \| \vec\theta/r_\mathrm{c} \|^2 \bigr)^{3/2}}
  \; .
\end{equation}
Only the central $3\arcmin$ of each profile were fitted with a
smoothed version of the model Eq.~\eqref{eq:iss}, with the same
smoothing length as the one used for the shear map (see Lombardi \&
Bertin 1998b for a discussion the relationship between the smoothing
scale used and the resulting mass map). The results of this fit are
shown in Fig.~\ref{fig29.ps} and summarized in Table~3. Note that the
total mass within $1$~Mpc obtained from the fit is roughly the double
of the mass assuming that the radial profile vanishes at large radii.
This discrepancy is related to the mass-sheet degeneracy and clearly
represents a serious limitation for a ``total'' mass estimate. We
stress in particular that the correction to be applied to the
``detected'' mass in order to obtain the total mass strongly depends on
the radial profile assumed for the cluster (the correction is exactly
$2$ for a singular isothermal sphere, see Eq.~\eqref{eq:1}).

\begin{table}[!b]
  \caption{Best fit parameters obtained in the B
    band. \label{tab:best-fit-par}}
  \small
  \begin{tabular}{l r l}
    \hline
    \hline
    Parameter              & \multicolumn{2}{c}{Value} \\
    \hline
    $r_\mathrm{c}$ (arcsec)&  $35$ & arcsec \\
    $\Sigma_0$             & $265$ & $\mbox{M}_\odot \ h_{50} \mbox{
      pc}^{-2}$ \\ 
    1D velocity dispersion & $850$ & $\mbox{km s}^{-1}$ \\
    Mass($r< 1$ Mpc) (fit) & $5.3 \times 10^{14}$ & $h_{50}^{-1}
    \mbox{ M}_\odot$ \\
    Mass($r <1$ Mpc) (direct) & $2.6 \times 10^{14}$ & $h_{50}^{-1}
    \mbox{ M}_\odot$ \\
    \hline
    \hline
  \end{tabular}
\end{table}

\subsubsection{Integrals of the shear}

An alternative method to evaluate the radial profile is to use
integrals of the shear (Fahlmann et al.\ 1994; Squires \& Kaiser
1996). In particular, for a point located at polar coordinates $(r;
\varphi)$ we define the \textit{tangential shear\/} as
$\gamma_\mathrm{t} = -( \gamma_1 \cos 2 \varphi + \gamma_2 \sin 2
\varphi)$. An estimator of the mass profile is given by
\begin{eqnarray}
  \zeta(\theta_1) &=& \langle \kappa \rangle(\theta_1) - \langle \kappa
  \rangle(\theta_1, \theta_2) \nonumber\\
  &=& \frac{1}{\pi(1 - \theta_1^2/\theta_2^2)}
  \int_{\theta_1}^{\theta_2} \!\!\diff r
  \int_0^{2\pi} \frac{\gamma_\mathrm{t}(r; \varphi)}{r} \,
  \diff\varphi \; ,
\end{eqnarray}
which measures the mean dimensionless mass density
$\langle\kappa\rangle(\theta_1)$ inside a disk of radius $\theta_1$
relative to $\langle\kappa\rangle(\theta_1, \theta_2)$, the mean in
an annulus of radii $\theta_1$ and $\theta_2$.

\begin{figure*}[!t]
  \centerline{\resizebox{\hsize}{!}{%
      \includegraphics{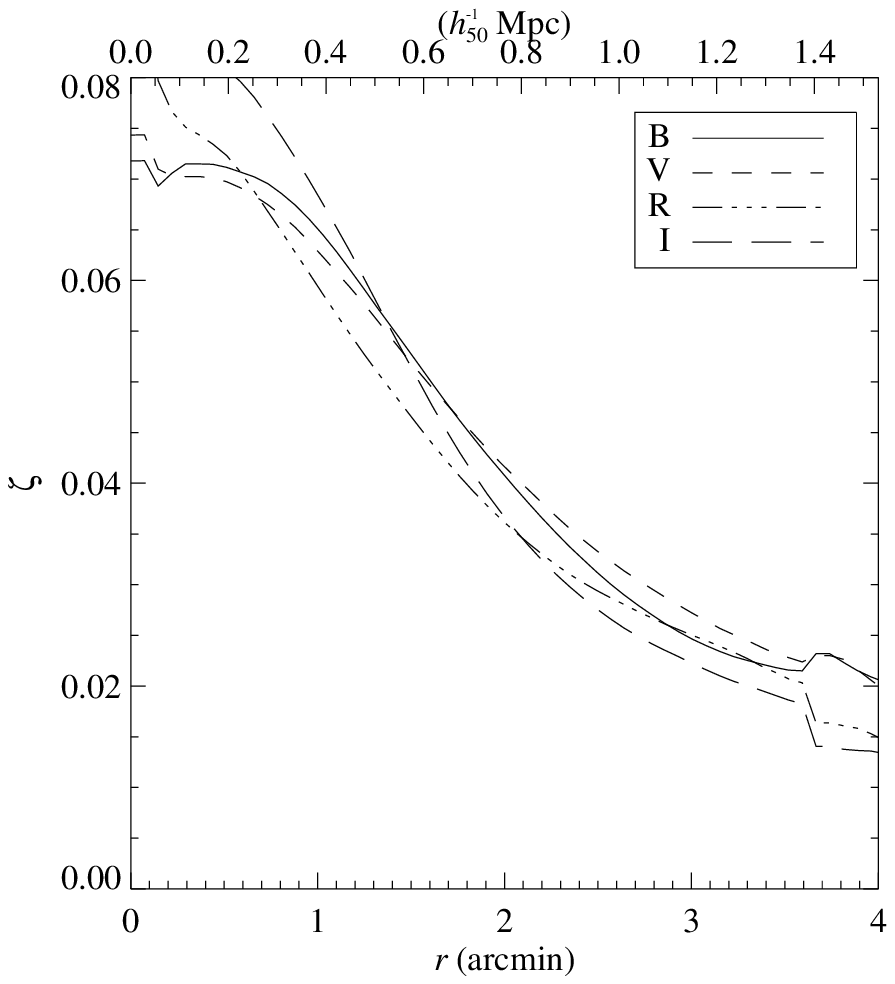}%
      \includegraphics{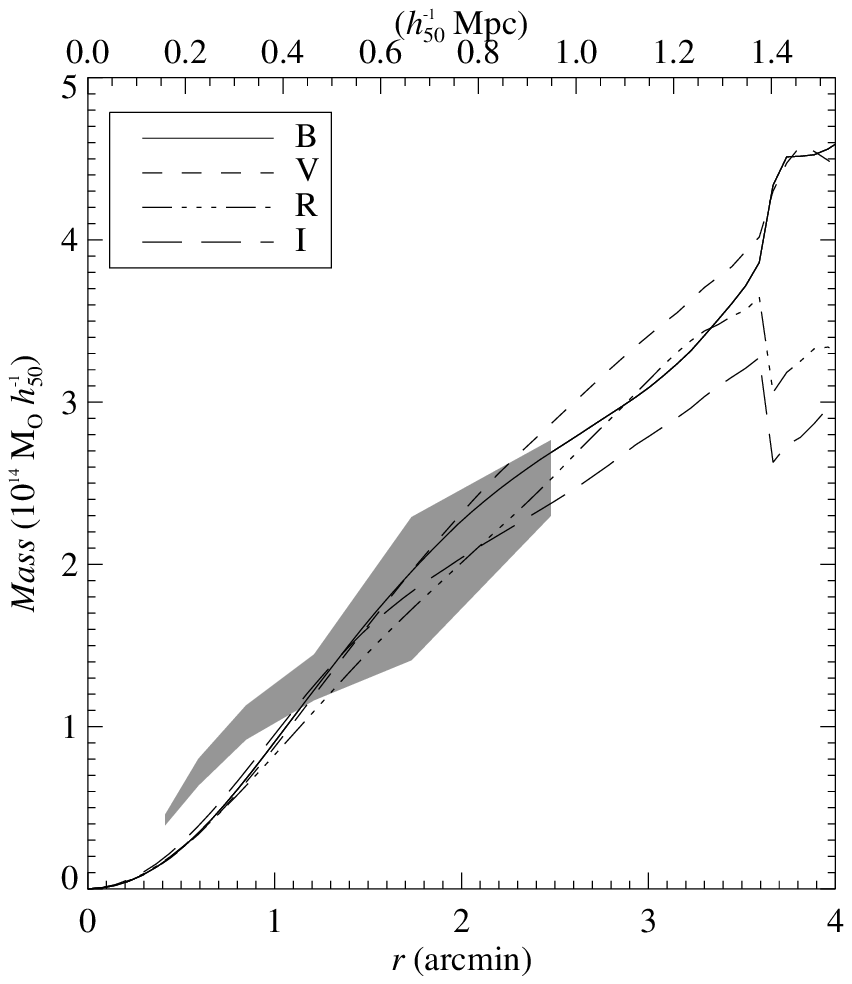}}}
  \caption{The (differential) mass profile estimator $\zeta$ (left) 
    and its integral, $Z=M(<r)$ (right), for the four passbands. The
    shaded are shows the result from the weak lensing analysis with a
    different method (see text). \label{fig25-26.ps}}
\end{figure*}  

We show $\zeta(\theta_1)$ in Fig.~\ref{fig25-26.ps} for each passband;
we used $\theta_2 = 3\arcmin\, 40\arcsec$ which defines the edge of
the images. The associated integral mass estimator, i.e.\ the quantity
\begin{equation}
  Z(\theta_1) = \Sigma_\mathrm{c} \pi \zeta(\theta_1, \theta_2) \theta_1^2
  \; ,
\end{equation}
is shown in Fig.~\ref{fig25-26.ps}.  In Fig.~\ref{fig25-26.ps} we also
show in gray the same estimate as obtained independently by one of us
(G.~Squires) using a slightly different technique (see Appendix).

\section{Mass profile from the virial analysis}
\label{sect:vir}

The analysis of the internal dynamics of MS1008, as traced by its
member galaxies, has already been discussed in the literature by the
CNOC collaboration.  The internal (line-of-sight) velocity dispersion
for this cluster was originally measured by Carlberg et al.\ (1996),
who found $\sigma_v = (1054 \pm 107) \mbox{ km s}^{-1}$, based on the
explicit background subtraction of the interloper contamination. More
recently, Borgani et al.\ (2000) consistently found $\sigma_v = (1042
\pm 110) \mbox{km s}^{-1}$, based on the completely different method
for removing interlopers, as described by Girardi et al.\ (1998, G98
hereafter). Carlberg et al.\ (1996) also estimated $M_{200} \simeq
2\cdot 10^{15} \mbox{ M}_\odot$ for the mass within the radius
$r_{200} \simeq 2.9 \,h_{50}^{-1}\mbox{ Mpc}$, which is defined as the
radius encompassing an average overdensity of $200$ (all the scales
from the original papers have been suitably converted to those
appropriate for the reference cosmological model assumed in our
analysis). Lewis et al.\ (1999) estimated the mass within a radius $r
\simeq 0.8\,h_{50}^{-1} \mbox{ Mpc}$ from \texttt{ROSAT} HRI X-ray
data and found $M_\mathrm{X} \simeq 3.6 \cdot 10^{14} h_{50}^{-1}
\mbox{ M}_\odot$. They also found this value to be consistent with
that obtained by extrapolating to this smaller radius the virial mass
estimate at $r_{200}$ with the NFW-like (Navarro et al.\ 1996)
profile, fitted by Carlberg et al.\ (1997) for the average density
profiles of member galaxies for CNOC clusters.

Here, we use the galaxy data (redshift and positions) as provided by
Yee et al.\ (1998) to determine the mass-profile of MS1008 from the
virial analysis. The procedure of the selection of member galaxies is
the same of G98. In addition, we also reject emission line galaxies,
since they are believed not to be fair tracers of the cluster internal
dynamics (e.g., Biviano et al.\ 1997). From G98 we also adopt the
procedure for estimating the velocity dispersion $\sigma_v$ and the
virial mass $M_\mathrm{vir}$, for which we provide here only a short
description. As for the velocity dispersion, we find
$\sigma_v=1033^{+115}_{-105}$ km s$^{-1}$, thus very close to previous
estimates.

In the case of a spherical system and under the assumption that mass
follows galaxy distribution, it is possible to show (e.g., Limber \&
Mathews 1960; The \& White 1986) that
\begin{equation}
  \label{eq:mvir}
  M_\mathrm{vir} = \frac{3 \pi}{2} \frac{\sigma_v^2 R_\mathrm{PV}}{G}-C
  \; .
\end{equation}
\noindent 
Here $C$ is a term of surface pressure correction, which takes into
account the fact that the radial component of the velocity dispersion
does not vanish within the observational aperture radius (Carlberg et
al. 1997). In the following, we assume a $20\%$ correction at the
virialization scale, which is typical for nearby and CNOC clusters
(G98; Carlberg et al.\ 1996).  The value of the
projected virial radius $R_\mathrm{PV}$ depends on the relative
(projected) positions of member galaxies within the sampled cluster
region. 

\begin{figure}[!t]
  \centerline{\resizebox{\hsize}{!}{\includegraphics{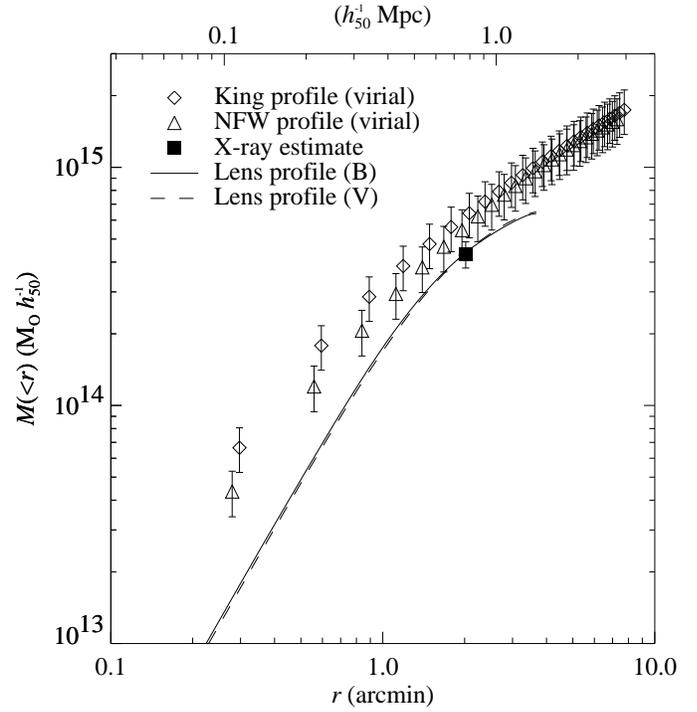}}}
  \caption{Comparison of different projected mass estimates.  Diamonds
  indicate results based on the virial mass from the present analysis,
  rescaled with the King-like density profile of Eq.~\eqref{eq:king}.
  Triangles are for the virial mass by Carlberg et al. (1996),
  rescaled with the NFW-like profile.  Errorbars are $1\sigma$
  uncertainties propagated from the statistical errors in the
  $\sigma_v$ estimate. The solid square is obtained by projecting the
  $X$-ray mass of Lewis et al.\ (1999), under the assumptions of
  spherical symmetry and NFW-like profile. The two curves are the
  mass profiles derived from weak lensing in two different bands.
  \label{fig:mass_vir}}
\end{figure}

Equation~\eqref{eq:mvir} provides a reliable estimate of the virial mass
only if galaxy redshifts are available over the whole virialized
region of the cluster.  Unfortunately this is not the case for MS1008
which is sampled only within about half of the virialized region (cf.\
Carlberg et al.  1996).  While this is not a serious problem for the
estimate of $\sigma_v$, whose profile flattens well within the
aperture radius (Borgani et al.\ 2000), it may represent a limitation
for $R_{PV}$, whose value increases with increasing clustercentric
distance. However, at each radius, an estimate of $R_\mathrm{PV}$ can
be obtained by assuming a shape for the density profile. For instance,
G98 estimated $R_\mathrm{PV}$ by assuming a King-like profile,
\begin{equation}
  \label{eq:king}
  \rho(r) \propto \frac{1}{\bigl[ 1 + (r / r_\mathrm{c})^2
  \bigr]^\alpha} \; ,
\end{equation}
with core radius $r_\mathrm{c} \simeq 0.1 \mbox{ Mpc}$ and shape
$\alpha \simeq 1.2$ [cf.\ their Eq.~(13)], which is the best fit to
the average galaxy density profile for an extended sample of local
clusters. They also verified this estimate of $R_{PV}$ to be reliable
for clusters sampled out to the virialization scale. In this way, we
find a total virial mass which fully agrees with that given by
Carlberg et al.\ (1996).  
{\mdf Diamonds}
in Fig.~\ref{fig:mass_vir} show the projected mass profile obtained
from the King-like profile of Eq.~\eqref{eq:king}.

Figure~\ref{fig:mass_vir} also shows with the 
{\mdf triangles}
the projected mass profiles obtained by scaling the original CNOC
estimate with the NFW-like profile fitted by Carlberg et al. (1997).
Since no definitive conclusion has been reached at present about the
most appropriate profile describing the cluster galaxy distribution
(see, e.g., Merritt \& Tremblay 1994; Adami et al.\ 1998), we prefer
here to show the results for both such profiles.  Althouh the two
virial mass determinations coincide on scales comparable to the
cluster virialized regions, the choice for the density profile affects
the mass reconstruction at small radii.

As a remarkable result, we find an almost perfect concordance between
our weak lensing mass and the $X$-ray mass by Lewis et al. (1999). 
{\mdf We stress that the weak lensing profiles shown in
  Fig.~\ref{fig:mass_vir} have been calculated by assuming a singular
  isothermal sphere for the cluster mass distribution (we recall that
  this assumption is important when removing the mass-sheet
  degeneracy).}
As for the comparison with the virial mass reconstruction, a fairly
good agreement is found at least for $r\magcir 0.4h_{50}^{-1}$ Mpc,
the virial mass being only slightly larger (at about 1$\sigma$
confidence level) than that provided by the weak lensing
reconstruction. The difference becomes significant only at smaller
radii, where, however, both the virial and the lensing masses become
more uncertain.

These results indicate that the weak lensing mass mass reconstruction
do provide results which agrees with other methods, based on the
assumption of dynamical equilibrium, at least in cases when
high-quality imaging data are available for a fairly relaxed cluster,
like MS1008 (see also Allen 1998).


\section{Mass to light ratio}

\begin{figure*}[!t]
  \centerline{\resizebox{\hsize}{!}{%
      \includegraphics{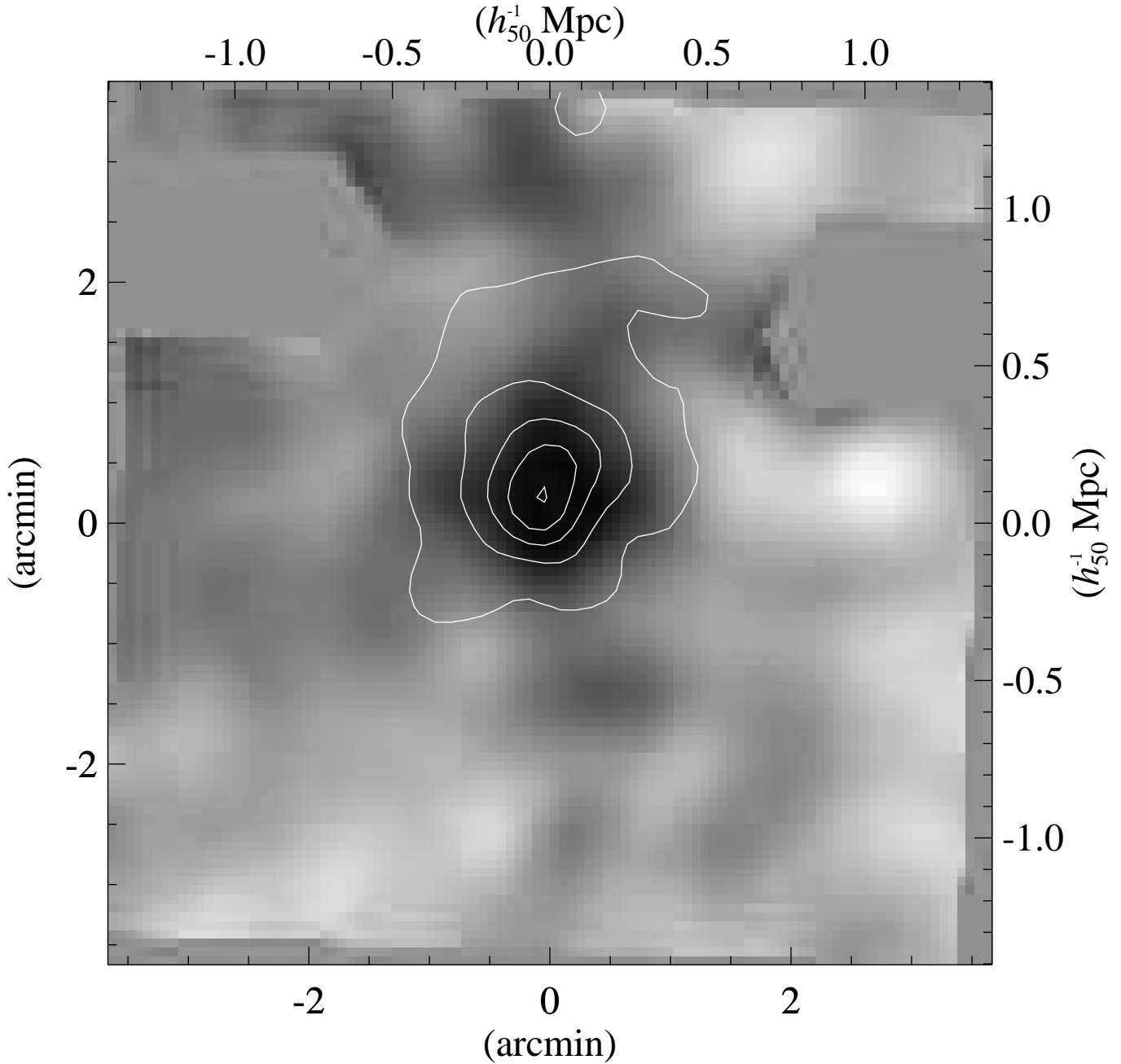}}}
  \caption{Left: Light distribution of MS1008 (gray scale) with
    overlaid mass contours, as in Fig.~\ref{fig:B-I}. 
    Dots represent galaxies selected as
    cluster members on the basis of their color (filled dots) or
    redshift (crossed dots). Right: The ROSAT HRI X-ray contours
    overlaid on the weak lensing mass distribution (gray scale).}
  \label{fig:mass-to-light}
\end{figure*}

To evaluate the mass to light ratio, we constructed a smoothed map of
the light distribution in the B band, using a Gaussian kernel with the
same scale as the one used for the shear map. The light density
distribution is shown in Fig.~\ref{fig:mass-to-light}, with overlaid
mass contours, and the cluster galaxies used in our analysis.  Galaxy
fluxes were measured with the \texttt{SExtractor} package (Bertin \&
Arnouts 1996), which ensures better object deblending in the cluster
core with respect to \texttt{IMCAT}.

Luminosities were computed by assuming a k-correction of $0.7$
magnitude in V band (Fukugita et al.\ 1995), appropriate for
ellipticals which dominate the cluster galaxy population. Luminosity
evolution of early type galaxies can be neglected at this low
redshift.  In our cosmological model the distance modulus is
$\mathit{DM} = 5 \log \bigl[ D(z_\mathrm{d}) (1 + z_\mathrm{d})^2 / 10
\mbox{ pc} \bigr] = 41.77 - 5 \log h_{50}$. The resulting rest-frame
V-band luminosity summed over all the cluster galaxies is
$L_\mathrm{V} = 5.6 \times 10^{12} h_{50}^{-2} \mbox{ L}_\odot$. This
estimate includes galaxies as faint as $M_* - 4$, hence the correction
for incompleteness should be negligible. Carlberg et al.\ found
$L_\mathrm{200} = 8.8 \times 10^{12} h_{50}^{-2} \mbox{ L}_\odot$
$(q_0 = 0.1)$ in the Gunn r band, at $r = R_\mathrm{200} \simeq 2.3
h_{50}^{-1} \mbox{ Mpc}$. Using $(V-R)_\mathrm{Ell} = 0.9$ and
$(r-R)_\mathrm{Ell} = 0.3$, our estimate translates to $L_\mathrm{Gunn
  r} \simeq 8.4 \times 10^{12} h_{50}^{-2} \mbox{ L}_\odot$ at $r
\simeq 1.4 h_{50}^{-1} \mbox{ Mpc}$ for $q_0=0.1$; hence it is in very
good agreement with $L_\mathrm{200}$ estimated by Carlberg et al.

Figure~\ref{fig:mass-to-light} clearly shows that, as commonly
observed in clusters (e.g.\ Clowe et al.\ 1998), the light roughly
traces the mass. The NE elongated morphology, clearly visible in the
smoothed light distribution, is also shared by the intra-cluster gas,
as visible in Fig.~\ref{fig:mass-to-light}, which shows the ROSAT-HRI
X-ray contours overlaid on the mass map (grey scale).  Interestingly,
the X-ray centroid is also displaced $15\arcsec$ North of the cD,
i.e.\ nearly coincident with the peak of the weak lensing signal.  The
X-ray HRI data are not deep enough to probe the outer regions of the
cluster. We also find some evidence for a drop of $M/L$ in the center
possibly due to the presence of the over-luminous cD galaxy.

In order to properly take into account the sheet degeneracy in the
evaluation of the mass-to-light ratio, we used an estimator similar to
$\zeta$ for the light distribution:
\begin{equation}
  \zeta_\mathrm{V}(\theta_1) = \langle L_\mathrm{V} \rangle(\theta_1)
  - \langle L_\mathrm{V} \rangle(\theta_2) \; ,
\end{equation}
where $\langle L_\mathrm{V} \rangle(\theta)$ is V band luminosity
averaged over a disk of radius $\theta$. Thus, $\zeta(\theta_1) /
\zeta_\mathrm{V}(\theta_1)$ yields an unbiased estimate of the mass-to-light
ratio (assuming that the light traces the mass).

Using this method we obtained $M/L_\mathrm{V} = 95 \, h_{50} \mbox{
  M}_\odot / \mbox{L}_\odot$ within a radius of $3'$, which compares
favorably with the value found by Hoekstra et al.\ (1998),
$M/L_\mathrm{V} = (90 \pm 13) h_{50} \mathrm{M}_\odot/\mbox{L}_\odot$,
for the CL~1358$+$62, with redshift ($z = 0.33$) and X-ray luminosity
($L_\mathrm{X} [0.3-3.5\, \rm{keV}] = 5.5 \times 10^{44} h_{50}^{-2}
\mbox{erg s}^{-1}$) very similar to MS1008.

\section{Conclusions}

We have studied the weak lensing mass distribution of the cluster MS1008
using multicolor imaging data obtained during the Science Verification
of FORS1 on the VLT. The depth, angular resolution, and image quality
across the entire field of view make these images a rare and ideal
dataset for weak lensing analysis. In addition, the information on
radial velocities of a large set of cluster galaxies provided by the
CNOC survey allows the mass of this cluster to be derived with dynamical
methods and compared with that reconstructed from the weak lensing.
These combined datasets allowed us to conduct a detailed study of the
systematic effects involved in cluster mass estimates.

We have found PSF distortions of FORS1 to be moderate and hence easily
removed with a low order polynomial fit. The B through I multicolor
information, as well as the redshift information from the CNOC
catalog, have allowed us to efficiently separate the background
galaxies from the cluster and foreground populations. Approximately
8000 objects have been detected, corresponding to $200 \mbox{ sources
  arcmin}^{-2}$. Further selection lead to approximatively $1700$
background galaxies with high signal-to-noise ratio which have been
used for the determination of lensing shear field. We have detected a
weak lensing signal out to $r \simeq 1.2 h_{50}^{-1}\mbox{ Mpc}$.  The
projected mass within $1 h^{-1}_{50} \mbox{ Mpc}$ is measured in the
range $(2.5\div 3.2)10^{14} h_{50}^{-1}\mbox{ M}_\odot$ in the four
filters, with typical statistical errors (deriving from uncertainties
in galaxy ellipticities) of 5\% .

We have discussed the impact of systematics in the weak lensing
reconstruction of the mass distribution in physical units, which can
be summarized as follows. The removal of the mass-sheet degeneracy is
inevitably model-dependent, inducing mass variations of $\sim\!  15\%$
at $r \approx 1 h^{-1}_{50} \mbox{ Mpc}$. [Note that the estimate of
this error is difficult and somewhat arbitrary since the error clearly
depend on the ``range'' of possible models allowed. As a result, the
estimate of the total error, given below, is also bound to be
inaccurate.  We stress, however, that a similar problem exists for
X-ray or virial mass estimate (cf., e.g., the different predictions
using King-like or NFW profiles if Fig.~\ref{fig:mass_vir}).]\@ By
polluting the background galaxy population with a significant fraction
of cluster galaxies, the total mass is biased low by $\mincir 5\%$.
However, different selections of background galaxies, as well as edge
effects due to masked portions of the images, do affect the morphology
of the shear maps.  We have not found any evidence for a substructure
of the cluster core; this in part is due to lack of resolution in the
mass reconstruction, set by the smoothing scale used.  However, we
have found the central region of the cluster to be elongated NS, with
some evidence of a displacement of the projected mass peak respect to
the cD galaxy, similarly to the X-ray emission.  Cosmic variance is
expected to affect the assumed redshift distribution of the background
field galaxies, and hence the derived cluster mass.  From a comparison
between the Hubble Deep Field North and South, cosmic variance would
introduce a systematic error of approximatively $10\%$ in the mass.
{\mdf This can also be taken as a rough estimate of the error due to
  cosmic variance.  Note also the main contribution to this error is
  from large scale structures rather than from Poisson noise in the
  HDF-N and in our field.  In other words, a simple error estimate
  which assumes that galaxies are uncorrelated would lead to an
  underestimate of the error.}
Altogether, the cluster light distribution traces the mass
distribution remarkably well. We have measured a mass-to-light ratio
of $M/L_\mathrm{V} = (95\pm 22) \, h_{50} \mbox{ M}_\odot /
\mbox{L}_\odot$ at radii $\ge 0.5 h_{50}^{-1}$ Mpc.

We have also compared the weak lensing mass profile with that derived
from a virial analysis of the CNOC redshift data, as well as with the
X-ray mass.  The mass derived from X-ray observations (Lewis et al.\ 
1999) is found in excellent agreement with our weak lensing
determination at $r \simeq 0.8 h^{-1}_{50} \mbox{ Mpc}$.  Different
approaches to estimate the cluster mass at the virialization scale
produce very similar results.  At scales $0.1 \,\mincir h_{50} r
\,\mincir 1$ Mpc, lensing and virial mass determinations can be
compared.  The virial mass reconstruction at these small radii
critically depends on the assumed density profile.  Using a King-like
profile describing the galaxy distribution of local clusters (G98; cf.
also Eq.~\eqref{eq:king}), we have obtained a mass which is more than
$2 \sigma$ away from the X-ray and weak-lensing mass. Assuming instead
a NFW profile, as fitted by Carlberg et al.\ (1997) to CNOC clusters,
a much better agreement is obtained over the entire overlapping
region.  Note that the discrepancy at small radii ($r < 1'$) between
lensing and virial estimate can be explained by recalling that the
lensing determination is affected by a Gaussian smoothing of $30''$,
as well as a departure from the weak lensing approximation. The former
effect leads to underestimate the mass in the core, while the latter
will generally bias the central mass high. Both effects clearly vanish
at large radii. In addition, we notice that the weak lensing data
alone cannot discriminate between different mass profiles (e.g.\ NFW
versus isothermal model).

This analysis shows that the combination of depth and good imaging
quality provided by VLT/FORS allows us to obtain high $S/N$ mass maps
via optimized weak lensing reconstruction methods. On the other hand,
these data have the virtue of revealing important systematic errors in
weak lensing analyses. We find that the mass-sheet degeneracy
dominates the budget of systematic errors and ultimately makes mass
measurements via weak lensing model dependent.

The results obtained here are, at a first look, in good agreement with
the mass reconstruction performed by Athreya et al.\ (2000). However,
a detailed analysis of their study reveals that a different recipe to
remove the mass-sheet degeneracy was used. Specifically, they set to
zero the mass at large radii ($3\farcm2$--$3\farcm6$). Had they used
our (model-dependent) method to break the mass-sheet degeneracy, their
estimate would have been a factor of two higher. Therefore, there is a
discrepancy of the same factor between the two analysis which, at
present, we are unable to explain.

\section{Acknowledgments}
The authors would like to thank the UT1 Science Verification team at
ESO for the preparation of this unique data set and its release to the
community.  We are particularly grateful to the VLT/UT1 Commissioning
Team for carrying out the observations.  We also thank Tom Broadhurst,
Tomas Erben, and Peter Schneider for useful discussions.  Support for
this work was provided by NASA through Hubble Fellowship Grant No.
HF-01114.01-98A from the Space Telescope Science Institute, which is
operated by the Association of Universities for Research in Astronomy,
Incorporated, under NASA Contract NAS5-26555.

\appendix

\section{Alternative Shear Analysis}

In the second analysis of the data, we followed the procedure used by
Hoekstra et al.\ (1998). The anisotropic part of the PSF is removed on
an object-by-object basis by correcting each object's ellipticity via 
\begin{equation} 
e_\alpha\to e_\alpha - {P_{\alpha\beta}^{\rm sm}\over
P_{\beta\beta}^{\rm sm}*}e_\beta*,
\end{equation}
where starred quantities refer to parameters measured for stellar
images. In this analysis, we calculated the stellar ellipticities and
polarizability employing a matched radius for the Gaussian weight
function as for the object in question. We fit a second order polynomial
to account for spatial variations in the PSF.

We next determine the shear from the anisotropy-corrected galaxy
polarizations. Both the (now effectively isotropic) PSF, and the
circular weight function, tend to make objects rounder. These effects
may be corrected for using the `pre-seeing shear polarizability'
\begin{equation}
P^\gamma = P^{sh} - P_*^{sh} / P_*^{sm} \times P^{sm}
\end{equation}
 introduced by Luppino \& Kaiser (1997), where the $*$ subscript again
refers to quantities determined from the objects classified as stars.

Operationally, to determine $\langle P^\gamma \rangle$, we split the
faint galaxy catalogs into six bins based on size $r_g$, with spacing
$dr_g = 0.25$ pixels. We reject galaxies with very large, small, or
negative values of $P^{sh} = (P^{sh}_{11} + P^{sh}_{22})/2$, and $P^{sm}
= (P^{sm}_{11} + P^{sm}_{22})/2$ (the off-diagonal terms typically being
very small). The rationale for this cut is that the calculation of the
shear and smear polarizabilities for these objects failed, typically
either because the objects themselves were very faint and near the
detection threshold, or due to the presence of very nearby object
contaminating the aperture over which the calculations were performed.

In each $r_g$ bin, we calculate the {\em median} galaxy polarizabilities,
$P^{sh}$ and $P^{sm}$, and reanalyze the stellar catalog, resetting the
value of the stellar $r_g$ to the bin size, and analyzing with a
Gaussian with scale $r_g$ and an aperture three times this radius.

\noindent We finally estimate the shear for each object using
\begin{equation} 
\gamma_\alpha=\frac{e_\alpha }{\langle P^\gamma \rangle}.
\end{equation}

\end{document}